\documentclass[aps,prb,twocolumn,notitlepage,nofootinbib]{revtex4-1}
\usepackage[T1]{fontenc} 

\usepackage[unicode,colorlinks]{hyperref}
\usepackage{mathtools}
\usepackage{amsmath,amssymb,amsfonts}
\usepackage{bbm}         
\usepackage{graphicx}
\usepackage{physics}

\usepackage{enumitem}
\setlist[itemize,1]{leftmargin=0cm}

\usepackage[dvipsnames]{xcolor}

\newcommand{\ignore}[1]{}

\newcommand{\Id}{\mathbbm{1}}
\renewcommand{\Im}{\mathop{\text{Im}}}
\renewcommand{\Re}{\mathop{\text{Re}}}

\renewcommand{\>}{\rangle}
\newcommand{\<}{\langle}

\DeclareMathOperator{\adj}{adj}

\begin{document}

\title{Exponentially growing bulk Green functions as signature of\\ nontrivial non-Hermitian winding number in one dimension}

\author{Heinrich-Gregor Zirnstein}
\affiliation{Institut f\"{u}r Theoretische Physik, Universit\"{a}t Leipzig, Br\"{u}derstrasse 16, 04103 Leipzig, Germany}
\author{Bernd Rosenow}
\affiliation{Institut f\"{u}r Theoretische Physik, Universit\"{a}t Leipzig, Br\"{u}derstrasse 16, 04103 Leipzig, Germany}

\date{July 14, 2020}

\begin{abstract}
\normalsize
A nonzero non-Hermitian winding number indicates that a gapped system is in a nontrivial topological class due to the non-Hermiticity of its Hamiltonian. While for Hermitian systems nontrivial topological quantum numbers are reflected by the existence of edge states, a nonzero non-Hermitian winding number impacts a system's bulk response. To establish this relation, we introduce the bulk Green function, which describes the response of a gapped system to an external perturbation on timescales where the induced excitations have not propagated to the boundary yet, and show that it will grow in space if the non-Hermitian winding number is nonzero. Such spatial growth explains why the response of non-Hermitian systems on longer timescales, where excitations have been reflected at the boundary repeatedly, may be highly sensitive to boundary conditions. This exponential sensitivity to boundary conditions explains the breakdown of the bulk-boundary correspondence in non-Hermitian systems: topological invariants computed for periodic boundary conditions no longer predict the presence or absence of boundary states for open boundary conditions.
\end{abstract}

\maketitle

%
\section{Introduction}

According to the topological classification of matter, two Hamiltonians belong to the same topological class if they can be continuously deformed into each other without closing a gap.~\cite{Hasan:2010a,Qi:2011,Bernevig:2013,Chiu:2016a,Bansil:2016} Quantities that do not change under such a deformation are called \emph{topological invariants}; Hamiltonians from the same class must yield the same value for the invariant. Of prime interest are topological invariants that can be related to physical observables; for example, the Chern number of a two-dimensional electronic insulator features directly in the Hall conductance.~\cite{Thouless:1982} The most general and striking consequence of the topological classification of insulators is the \emph{bulk-boundary correspondence}, which predicts that if the topological invariants calculated from the Bloch Hamiltonian of a periodic system are nonzero, then boundary eigenstates exist in a system with open boundary conditions.~\cite{Halperin:1982,Hatsugai:1993,Hasan:2010a,Essin:2011,Graf:2013,Avila:2013} This holds if the system Hamiltonian is Hermitian, which means that it conserves the number of particles.

Non-Hermitian Hamiltonians~\cite{Bender:1998,Rotter:2009,Cao:2015,Zhen:2015} describe open systems with loss and gain, where probability (quantum mechanical case) or energy (classical case) are no longer conserved~\cite{Longhi:2017,El-Ganainy:2018}.
For such systems, generalized topological invariants can be defined~\cite{Bergholtz:2019a,Gong:2018,Kawabata:2019b,Zhou:2019a,Shnerb:1998,Lee:2016,Leykam:2017,Shen:2018,Lieu:2018,Hirsbrunner:2019,Longhi:2019a,Chen:2018c,Wanjura:2020}, which have no counterpart (i.e.~vanish) in Hermitian systems, for example the so-called non-Hermitian winding number~\cite{Gong:2018,Lee:2016,Leykam:2017}.
The physical consequences of these novel invariants are not fully understood yet. Unlike in the Hermitian case, the bulk-boundary correspondence seems to break down: Topological invariants calculated from the energy bands for periodic boundary conditions no longer fully predict the presence of zero energy eigenstates in systems with open boundary conditions~\cite{Xiong:2018,Kunst:2018,Yao:2018b,Lee:2019c,Jin:2019a,Herviou:2019,Ge:2019,Zirnstein:2019,Borgnia:2020,Brzezicki:2019a,Yokomizo:2019,Kawabata:2020,Yang:2019g}.

In this work, we establish an experimentally observable implication of the non-Hermitian winding number in one-dimensional lattice systems. While we do not provide a final answer to the question of the bulk-boundary correspondence for non-Hermitian systems, we identify a key mechanism for its breakdown: We introduce the \emph{bulk regime}, which captures the response of a system to an external perturbation on  timescales where the induced bulk excitations have not propagated to the boundary yet, and boundary conditions do not yet influence the response. Our main result is to show that the response in the bulk regime will \emph{grow} in space if the non-Hermitian winding number is nonzero.
This is in strong contrast to the response of a system with periodic boundary conditions, which always decays in space, and where the response is considered only after excitations had sufficient time to propagate through the entire system repeatedly.
The spatial growth of the bulk response makes a high sensitivity to boundary conditions plausible, and also explains the subsequent failure of topological invariants calculated for periodic systems to fully capture the response of  systems with open boundary conditions: such topological invariants  no longer predict the presence or absence of boundary states.
Conversely, we establish that a spatial growth of the response in the bulk regime does not occur for systems with a line gap in the energy spectrum, making it plausible that the bulk-boundary correspondence established for Hermitian Hamiltonians continues to hold in this case. This work complements our results in Ref.~\cite{Zirnstein:2019} by providing a precise definition of the bulk Green function and proving spatial growth for general lattice systems. We note that our analysis of the bulk Green function is more general than a description of the non-Hermitian skin effect~\cite{Xiong:2018,Yao:2018b,MartinezAlvarez:2018,MartinezAlvarez:2018a,Zhang:2020e,Okuma:2020,Okuma:2020a,Xiao:2020,Helbig:2020,Hofmann:2020,Yoshida:2020,Longhi:2019d}, which is a property of eigenstates for specific boundary conditions, while the bulk Green function is not affected by boundary conditions at all.

To describe a spatially growing response function, we consider the Bloch Hamiltonian $H(e^{ik})$ not only for real momenta $k$, but will generalize to a complex variable, and consider $H(z)$. The response function (Green function) is essentially an inverse of the Hamiltonian, $G(E) \sim [E-H(z)]^{-1}$, which admits a partial fraction decomposition as a sum over terms $[z-z_j(E)]^{-1}$, where the $z_j(E)$ are complex roots of the determinant of the Hamiltonian. This allows us to derive explicit formulas for both the periodic Green function and the bulk Green function, which are represented as sums of powers $[z_j(E)]^x$, where $x$ is an integer denoting the spatial distance in units of the lattice contant between the spatial point where the system is excited and the point where the response is measured. To achieve our goal of describing the response in the bulk regime, we introduce a definition of the bulk Green function via analytic continuation, which proceeds by adding a uniform dissipation $\Gamma$ large enough to suppress any amplification, and then following the path of the poles $z_j(\Gamma)$ in the complex plane back to $\Gamma=0$. If, at the end of this procedure, lattice sites with positive $x$ receive contributions from roots with $|z_j|>1$, then the response grows as the distance is increased. This happens precisely if at least one root has moved from inside the unit circle to the outside or vice versa. But a nonzero net movement can be detected by the non-Hermitian winding number, i.e.~the winding number of the determinant $\det(H(z))$, which for this reason indicates the exponential growth of the bulk Green function.

The paper is organized as follows:
In Section~\ref{sec:bulk}, we define the bulk regime and motivate the definition of the bulk Green function, which describes  the response of a system on  timescales where induced excitations have not propagated to the boundary yet.
In Section~\ref{sec:bulk-green}, we compute the periodic Green function in terms of the roots $z_j$ of the determinant of the Hamiltonian, define the bulk Green function via analytic continuation, and determine the circumstances under which the bulk Green function differs from the periodic one.
In Section~\ref{sec:winding}, we define the non-Hermitian winding number for one-dimensional systems, show that it counts the difference between roots $z_j$ inside and outside the unit circle, and establish that a nonzero winding implies that the bulk Green function exponentially grows in space.
In Section~\ref{sec:conclusion}, we present our conclusions.
In Appendix~\ref{sec:general-hopping}, we relax an assumption on the Hamiltonian which was made in Section~\ref{sec:bulk-green} for the sake of clarity. 
In Appendix~\ref{sec:schroedinger}, we justify the definition of the bulk Green function via analytic continuation by showing that it indeed agrees with the long-time limit of the driven Schrödinger equation whenever this limit exists.
In Appendix~\ref{sec:green-decay}, we present an argument that the solution to the driven Schr\"{o}dinger function, and thus the bulk Green function, decays in space for so-called purely dissipative Hamiltonians. 
In Appendix~\ref{sec:long-times}, we discuss a lattice model for which the driven Schr\"{o}dinger equation can be solved exactly.
In Appendix~\ref{sec:green-growth}, we give an example of a non-Hermitian Hamiltonian whose bulk Green function grows in space, but which can be deformed into a Hermitian Hamiltonian with zero non-Hermitian winding number without closing the point gap, thus demonstrating that an exponentially growing bulk response does not imply a nonzero winding number.

%
\section{Bulk regime}
\label{sec:bulk}

In this section, we define the \emph{bulk regime}, which characterizes the response of a system on  timescales where induced excitations have not propagated to the boundary yet, and where boundary conditions do not  yet influence the response. We refer to the Green function in this regime as \emph{bulk Green function}. This is in contrast to the response of a system with periodic boundary conditions, for which topological invariants are traditionally computed, where excitations had time to repeatedly propagate through the entire system, and the periodic boundary conditions may influence the response. For non-Hermitian systems, these two responses may differ significantly: The response in the bulk regime may grow in space, which implies that the response on longer timescales will be highly sensitive to boundary conditions.

We consider one-dimensional systems with translational invariance and finite range hopping, described by a Hamiltonian operator $\hat H$ acting on wave functions $\psi(x)$ with internal degrees of freedom. The discussion in this section applies to both continuum models and lattice models, while we will consider more specific lattice models in subsequent sections.

To physically motivate the Green function in the bulk regime, we appeal to the known fact that Green functions describe solutions to the equation of motion in response to periodic driving. In particular, we consider the Schr\"o{}dinger equation for a wave function $\psi(t,x)$ subject to a driving force at position $y=0$ that oscillates with frequency (energy) $E$:
\begin{subequations}
\label{eq:driven-schroedinger}
\begin{align}
    i\partial _t \psi (t,x)
    &= \hat H \psi (t,x) - \psi _0e^{-iEt}\delta_{x,0}
    \\
    \psi (0,x) &= 0
.\end{align}
\end{subequations}
Here, $\hat H$ is the non-Hermitian Hamiltonian operator of interest and $\psi_0$ is a vector that records the amplitudes of the drive.
We now discuss the time evolution of the solution qualitatively. Initially, the wave function vanishes, but the driving force will inject nonzero amplitude into the system. Under the assumption of short-range hopping in $\hat{H}$, an excitation initially  localized at $x=0$ will propagate with a maximum velocity, say $v$. For a system of length $L$, the excitation will not reach the boundary until a time $T \sim L/v$. Meanwhile, the driving will continue to inject excitations at position $x=0$. As a result of this continuous injection, we expect that the wave function near the origin will settle into a stationary shape, reflecting the steady flow of excitations that are added to this region and propagate away from it. We will identify the bulk Green function $G_{\text{bulk}}(E;x)$ with this wave shape. Later, at times $t \gtrsim 2T$, the excitations that were injected previously will have had time to reach the boundary, be reflected, and propagate back to the origin, eventually disturbing the initial stationary state.

\begin{figure}
\includegraphics{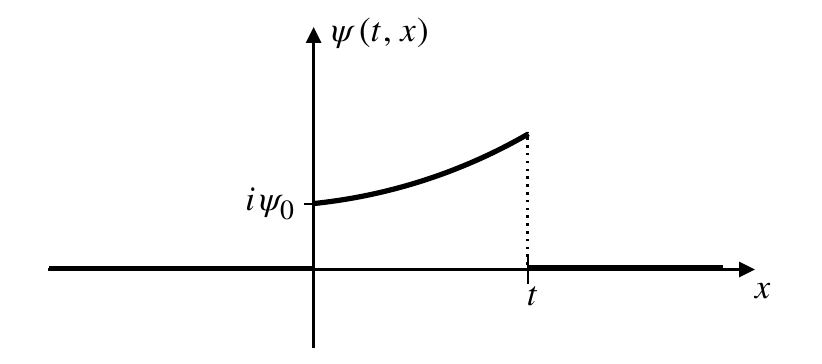}
\caption{%
Illustration of the bulk regime: The solution $\psi(t,x)$ to the driven Schr\"{o}dinger equation for the Hamiltonian Eq.~\eqref{eq:continuum-long-time} grows exponentially in space  as the induced excitations travel to the right (for $\gamma > 0$). Nonetheless, the wave function converges to a steady state pointwise for every position $x$  in the limit $t\to\infty$.
\label{fig:bulk_regime}}%
\end{figure}

More formally, the Green function in the bulk regime is given by the solution to Eq.~\eqref{eq:driven-schroedinger} in the simultaneous limit where both $t\to \infty$ and $L\to \infty$, but the system size $L$ grows asymptotically faster than the maximum propagation distance $vt$, such that $vt/L \to 0$. Mathematically, this is conveniently accomplished by letting the spatial domain of definition of $\psi(t,x)$ be the infinite real axis (or lattice), and then taking the pointwise limit $\lim_{t\to\infty} \psi(t,x)e^{iEt}$. In order to obtain a well-defined limit, we compensate the oscillatory behavior of the wave function by including the  factor $e^{iEt}$.
We illustrate this concept with the help of a specific example, the continuum model
\begin{equation}
\label{eq:continuum-long-time}
    \hat H = -i\partial _x + i\gamma  \ 
\,.\end{equation}
This Hamiltonian describes a wave traveling to the right with an additional non-Hermitian term. The real parameter $\gamma $ may be negative (loss) or positive (gain).
For the case $E=0$, the solution to the driven Schr\"o{}dinger equation \eqref{eq:driven-schroedinger} is readily calculated as
\begin{equation}
\label{eq:continuum-solution}
    \psi (t,x) = i\psi _0 [\theta (x)-\theta (x-t)]e^{\gamma x} \ 
,\end{equation}
where $\theta $ denotes the Heaviside theta function. The wave function is nonzero only in the spatial interval $0 < x < t$ and has already reached a stationary state in this interval [see Fig.~\ref{fig:bulk_regime}]. Thus, we can straightforwardly take the limit $t\to \infty$, pointwise for each position, and obtain the bulk Green function
\begin{align}
    G_{\text{bulk}}(0;x)
    &:= \lim_{t\to \infty } \psi (t,x)/\psi _0
    = i\theta (x) e^{\gamma x} \ \ 
\label{eq-green-gain}
.\end{align}
For $\gamma$ negative (loss), the Green function decays in space as we would expect in a gapped Hermitian system for excitation energies smaller than the energy gap, but for $\gamma$ positive (gain), it grows exponentially. Compared to Hermitian systems, the long-time limit only exists in the sense of pointwise convergence, but not in the sense of convergence in the Hilbert space norm, because the latter is only defined for wave functions that decay at spatial infinity. This implies that we will have to take some care when applying mathematical techniques known from the Hermitian case, because they typically assume that wave functions have a finite Hilbert space norm.

In the following section, we will derive an explicit expressions for the bulk Green function of general non-Hermitian lattice systems. However, we will not compute the long-time limit $t\to\infty$ directly, but will instead perform an analytic continuation of the periodic Green function with additional dissipation. The connection between these two approaches can be established by considering the Laplace transform of the Green function, which we describe in detail in Appendix~\ref{sec:schroedinger}. In fact, we will formally define the bulk Green function as the analytic continuation Eq.~\eqref{eq:bulk}. This definition is more general than the pointwise limit $t\to\infty$, because it also applies to situations where no stationary state is reached: In some non-Hermitian systems, it may happen that the wave function is amplified faster than excitations can propagate away, and the solution to the driven Schr\"{o}dinger at $x=0$ grows indefinitely in time. The physical interpretation of the analytic continuation in this case is outside the scope of this work however.

%
\section{Periodic versus bulk Green function}
\label{sec:bulk-green}

In this section, we derive derive explicit expressions for the periodic and the bulk Green function, and show that in non-Hermitian systems the two Green functions no longer need to agree with each other in the limit of large systems size, contrary to the Hermitian case. However, we will demonstrate that periodic and bulk Green function agree with each other in two important cases: in (i) so-called purely dissipative systems, and (ii) in systems with a real line gap~\cite{Kawabata:2019b}, where the imaginary axis is excluded from the spectrum of the Bloch Hamiltonian. This makes it plausible that the bulk-boundary correspondence continues to hold in both these cases, which implies that topological invariants computed for periodic boundary conditions predict boundary states in a system with open boundary conditions.

We consider one-dimensional lattice systems with translational invariance and finite range hopping. By suitably enlarging the unit cell, we can describe them by a Hamiltonian operator $\hat H$ with nearest-neighbor hopping only. Its action on a lattice wave function $\psi(x)$ is defined by $(\hat H\psi)(x) = H_{-1}\psi(x-1) + H_0\psi(x) + H_1\psi(x+1)$, where $H_{-1},H_0,H_1$ are complex $N\times N$-matrices, and $N$ denotes the number of sites (degrees of freedom) within a unit cell. We have chosen to label lattice sites by an integer $x$ and set the lattice spacing to unity.
The well-known Bloch Hamiltonian $H(e^{ik})$ is obtained by applying the Hamiltonian operator to plane waves $\psi(x) = e^{ikx}$ with real momentum $k$. However, to compute Green functions, it will be very useful to consider its analytic continuation to complex momenta by replacing the phase increment between adjacent lattice sites $e^{i k }$ with a complex number $z$. In this way, we are led to consider the matrix Hamiltonian
\begin{equation}
  \label{eq:hamiltonian}
  H(z) = H_{-1} z^{-1} + H_0 + H_1 z
,\end{equation}
which we simply call the \emph{Hamiltonian}, for any complex number $z$, and not just for pure phases $e^{ik}$. This corresponds to considering the action on wave functions of the form $\psi(x)=z^x$, which describe exponentially damped or growing waves. A Hermitian Hamiltonian is characterized by the requirements $H_{-1}=H_1^\dagger$ and $H_0=H_0^\dagger$, but there are no such constraints for non-Hermitian Hamiltonians.

We consider systems with an energy gap: A Hamiltonian is said to have a \emph{point gap}~\cite{Kawabata:2019b} at zero energy if none of the (energy) eigenvalues of the Bloch Hamiltonian $H(e^{ik})$ is equal to zero for any real momentum $k$. This is the same as saying that the matrix $H(z)$ is invertible on the unit circle $|z|=1$. Since the Green function at zero energy is an integral over the inverse of the Hamiltonian, this condition ensures that the Green function is well-defined. We emphasize that no condition is imposed away from the unit circle, $|z|\neq 1$. While this notion of energy gap is sufficient for Hermitian systems, for non-Hermitian systems, another notion was introduced in the literature~\cite{Kawabata:2019b}, called \emph{real line gap}, which is stronger and requires that the eigenvalues of $H(e^{ik})$ need to avoid not just zero, $E\neq0$, but the entire imaginary axis, $\Re(E)\neq 0$. Later, we will show that this stronger condition implies that bulk and periodic Green function agree with each other asymptotically; but for now, we only assume a point gap.

\begin{figure}
\includegraphics{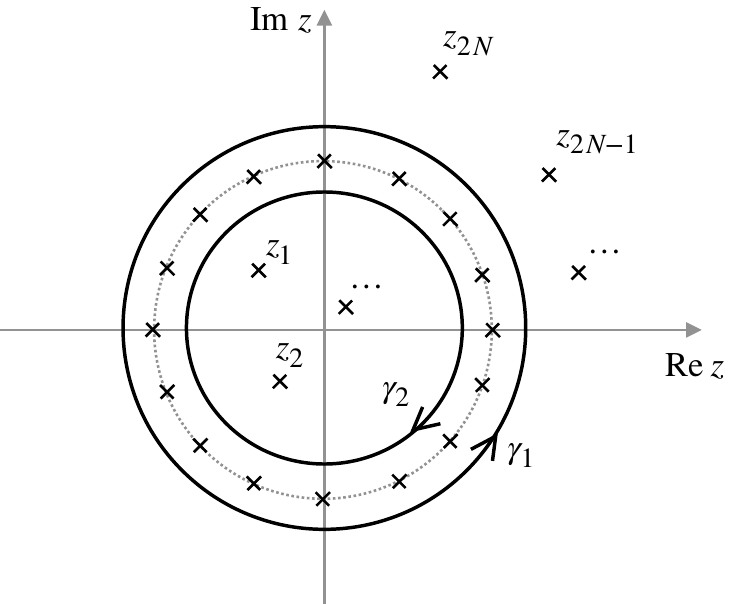}
\caption{%
Illustration of the integration contour for the integral representation of the periodic Green function $G_{\text{period}}$, Eq.~\eqref{eq:period-contour}, consisting of two paths $\gamma_1$,$\gamma_2$ that form the boundary of a thin annulus that encloses the unit circle. The poles of the integrand are indicated by crosses: The factor $1/(z^L-1)$ contributes poles on the unit circle, while the inverse of the Hamiltonian contributes poles $z_1,\dots,z_{2N}$ that lie outside the annulus.
\label{fig:roots_periodic}}%
\end{figure}

We now discuss the Green function for periodic boundary conditions, or \emph{periodic Green function} $G_{\text{period}}(E)$ in more detail.
It represents the resolvent operator $[E-\hat H]^{-1}$ for a periodic system.
In the following, we focus on zero energy, $E=0$. Periodic boundary conditions allow us to represent the Green function as a discrete Fourier series
\begin{equation}
  \label{eq:period}
  G_{\text{period}}(0;x)
  = -\frac1L \sum_{m = 0}^{L-1} \frac{e^{i k_m x}}{H(e^{i k_m})}
  ,\quad k_m=\frac{2\pi m}{L}
,\end{equation}
where $L$ is the number of sites in the lattice. Since we have assumed a point gap, the matrix inverse $1/H(e^{ik_m})$ is well-defined.
We can bring this expression into a more explicit form by using the residue theorem and rewriting it as a contour integral. In particular, the exponentials $e^{ik_m}$ are the poles of the function $g(z) = 1/[z(z^L-1)]$ on the unit circle, with residues equal to $1/L$. Thus, we find
\begin{align}
\label{eq:period-contour}
    G_{\text{period}}&(0;x)
    =
    -\frac1{2\pi i} \oint_\gamma \frac{dz}z \frac1{H(z)} \frac{z^x}{z^L - 1} \ 
\, ,\end{align}
where the integration contour $\gamma = \gamma _1 \cup \gamma _2$ is the boundary of an thin annulus that encloses the unit circle, e.g.~$\gamma _{1,2}(t) = (1 \pm \varepsilon )e^{\pm 2\pi it}$, as illustrated in Fig.~\ref{fig:roots_periodic}.
To proceed, we will use the residue theorem a second time, but we will now evaluate Eq.~\eqref{eq:period-contour} as a sum over all poles that lie outside the annulus. These poles are due to the inverse of the Hamiltonian.

A compact way to employ the residue theorem as described above is to utilize a partial fraction decomposition for the inverse Hamiltonian, which we will derive in the following form:
\begin{equation}
  \label{eq:partial-fraction}
  -\frac1{z} \frac{1}{H(z)} = \sum_{j=1}^{2N} G_j \frac1{z-z_j}
\, ,\end{equation}
where $G_1, \dots, G_{2N}$ are complex $N\times N$ matrices, and $z_1, \dots, z_{2N}$ are complex numbers. These number are the zeros of the determinant $\det(H(z))$, and for a Hamiltonian with a point gap, they do not lie on the unit circle, $|z_j|\neq 1$. For this decomposition to hold, we need to assume: (i) that the roots $z_j$ are distinct from each other, and (ii) that the hopping matrices $H_{-1}$ and $H_{1}$ in the Hamiltonian \eqref{eq:hamiltonian} are invertible.
The first assumption (i) is mild: It is true for a generic non-Hermitian Hamiltonians with a point or line gap, and the rare case where two or more roots coincide will either not affect the gist of our results or correspond to a transition point.
The second assumption (ii) is more restrictive, it implies that the hopping between neighboring unit cells involves all internal degrees of freedom of these cells. However, we make this assumption only for the sake of simplicity, both the results and the main ideas of our argument hold without it; in fact, we present the details for the general case in Appendix~\ref{sec:general-hopping}.
To derive the partial fraction decomposition~\eqref{eq:partial-fraction}, the first step is to use that the inverse of a matrix can be expressed in terms of its determinant and its adjugate matrix (Cramer's rule):
\begin{equation}
    \label{eq:adjugate}
    \frac1{z} \frac1{H(z)} = \frac{1}{\det(z H(z))}\adj(z H(z)) \ 
.\end{equation}
We now argue that both the adjugate matrix and the determinant are polynomials in the matrix entries, and thus polynomials in $z$, because the entries of the matrix $zH(z)$ are themselves polynomials in $z$.
Explicitly, the determinant can be expressed as
\begin{equation}
  \label{eq:determinant-polynomial}
  \det(zH(z)) = \sum_{m=0}^{2N} a_{m} z^m \ 
,\end{equation}
with complex coefficients $a_0,\dots,a_{2N}$. Using the definition of  the determinant as an alternating sum over permutations, one can see that the lowest and highest coefficients are $a_0 = \det(H_{-1})$ and $a_{2N} = \det(H_1)$, which are nonzero by assumption (ii). Thus, this polynomial has degree $2N$ and nonzero roots $z_1,\dots,z_{2N}$. They allow us to express the determinant of the Hamiltonian as a product
\begin{equation}
\label{eq:determinant-factorization}
  \det(H(z)) = \det(H_1)z^{-N} (z-z_1)\cdots(z-z_{2N}) \ 
.\end{equation}
Here, we have used the scaling identity $\det(zH(z))=z^N\det(H(z))$.
Then, performing a partial fraction decomposition for each matrix entry of the right-hand side in Eq.~\eqref{eq:adjugate} gives the result Eq.~\eqref{eq:partial-fraction}.
Assumption (i) implies that the roots are distinct; thus each pole is simple, and no higher powers appear in the decomposition.
Moreover, there is no constant polynomial term in the decomposition~\eqref{eq:partial-fraction}, because our assumption $\det(H_1)\neq 0$ implies that the left-hand side tends to zero as $z\to \infty$. Likewise, $\det(H_{-1})\neq 0$ guarantees that there is no pole at $z=0$.

We now evaluate the periodic Green function by performing the contour integral Eq.~\eqref{eq:period-contour} over the complement of the annulus.
For this purpose, we assume that the values of $x$ lie in the range $0,1,\dots,L-1$ and continue periodically later on. Then, the integrand has no poles at $z=0$ or at $z=\infty$, and we can focus on the poles at the points $z_j$. Integration over the inner contour $\gamma_2$ will have contributions due to roots with $|z_j|<1$, while applying the residue theorem to the outer curve $\gamma_1$ on the Riemann sphere will yield contributions due to roots with $|z_j|>1$. Taken together, we find
\begin{equation}
  \label{eq:period-explicit}
  G_{\text{period}}(0;x) = - \sum_{j=1}^{2N} G_j \frac{(z_j)^{x}}{(z_j)^L - 1}
  \, ,
  \text{ for } x=0,1,\dots,L-1
.\end{equation}
This expressions simplifies further if we take the limit of an infinite system, $L\to \infty$. For this, we need to label the positions in a symmetric way by using the spatial periodicity of $G_{\text{period}}$, such that the range of site labels becomes $x=-L/2,\dots,-1,0,1,\dots,L/2$, which turns into the set of integers in the infinite system limit. The limit is straightforward for positive labels, but for negative labels, we need to keep in mind that Eq.~\eqref{eq:period-explicit} was originally derived for positive labels, so we have to use periodicity, $x\equiv L+x$, to rewrite the summands as $(z_j)^{x}/((z_j)^L-1) = (z_j)^{L+x}/(1-(z_j)^{-L})$, and keep $L+x$ fixed when performing the limit. In this way, we obtain
\begin{equation}
\label{eq:period-limit}
    G_{\text{period}}(0;x) \stackrel{L\to \infty}{=}
    \begin{cases}
        \sum\limits_{|z_j|<1} G_j (z_j)^{x}\ , &\!\!\!\!\text{if } x = 0,1,\dots  \\
        \sum\limits_{|z_j|>1} (-1)G_j (z_j)^{x} \ , &\!\!\!\!\text{if } x = 0,-1,\dots
\end{cases}\end{equation}
Above, we have used that the factor $1/([z_j]^L-1)$ asymptotically tends to $(-1)$ when the root lies inside the unit circle, $|z_j|<1$, and vanishes when the root lies outside, $|z_j|>1$.
Physically, the Green function describes the response of the system when it is perturbed at the center position $x=0$. Thus, the response to the left of the perturbation is found at negative $x$, while the response to the right is found at positive $x$. We see that in each case, a different set of roots contributes to the Green function. Moreover, the contributions are such that each term decays as we move away from the perturbation, $x\to\pm\infty$.

In the following, we will  not assume periodic boundary conditions anymore, and turn our discussion to the bulk regime as introduced in Section~\ref{sec:bulk}.
In a Hermitian system, one would compute the bulk Green function from the periodic Green function by taking the limit of the system size going to infinity, obtaining Eq.~\eqref{eq:period-limit}. However, and this is a key result of our work, this procedure will fail to model the bulk regime for general non-Hermitian Hamiltonians, because it assumes that imposing periodic boundary conditions before taking the infinite system limit commutes with the long-time limit.
Before discussing this general case, we will first define a subclass of Hamiltonians, which we will call ``purely dissipative'' Hamiltonians, where the bulk Green function can still be obtained from the periodic one.
To establish this connection, we appeal to the analyticity properties of the retarded Green function for a Hermitian Hamiltonian, in particular to the fact that it is a complex analytic function of the energy parameter $E$ in the upper half-plane. In addition, due to the fact that the Green function is an inverse, the poles of the Green function are the eigenvalues of the Bloch Hamiltonian. Thus, if the eigenvalues of a general non-Hermitian Bloch Hamiltonian all lie in the lower half-plane, then it is plausible that the infinite system limit of the periodic Green function describes a retarded response on short to medium timescales, i.e.\ in the bulk regime.
For such Hamiltonians, any excitation dissipates because all eigenvalues have negative imaginary parts. However, due to the fact that the eigenvectors of non-Hermitian Hamiltonians need not be orthogonal, we introduce a slightly stronger notion of dissipation: We call a Hamiltonian \emph{purely dissipative} if the matrix corresponding to its antihermitian part is always negative semidefinite,
\begin{equation}
  \label{eq:purely-dissipative}
  (i/2)[H(e^{ik})^\dagger - H(e^{ik})] \leq 0 \text{ for every } k\in[-\pi,\pi]
.\end{equation}
This condition ensures that the total probability of any wave function decreases monotonically, whereas the requirement that all eigenvalues have negative imaginary part may only imply that decay will happen eventually (see also Eq.~\eqref{eq:norm-decrease} in the Appendix).
This example motivates us to consider, for any general non-Hermitian Hamiltonian $H(z)$, the family of Hamiltonians obtained by adding an auxiliary, uniform dissipation,
\begin{equation}
  \label{eq:hamiltonian-gamma}
  H_{\Gamma}(z) := H(z) - i\Gamma\Id
,\end{equation}
as this will shift the energy eigenvalues towards the lower half-plane.
If we choose $\Gamma$ slightly larger than the maximum gain,
\begin{equation}
  \label{eq:gamma-maximum-gain}
  \Gamma_0 = \max_{k}\{\text{gain}(H(e^{ik}))\} + \varepsilon
  ,\quad \varepsilon > 0
,\end{equation}
then all eigenvalues of $H_{\Gamma_0}(z)$ are moved far into the lower half-plane, and the shifted Hamiltonian $H_{\Gamma_0}(z)$ becomes purely dissipative. Here, $\text{gain}(H)$ denotes the maximum eigenvalue of the Hermitian matrix $(i/2)(H^\dagger - H)$.

To find the bulk Green function of a general non-Hermitian Hamiltonian, we now define an auxiliary Green function for the Hamiltonians $H_\Gamma(z)$ via the Fourier integral
\begin{equation}
  \label{eq:green}
G(i\Gamma;x)
  :=
- \int_{-\pi}^\pi \frac{dk}{2\pi}\, \frac{e^{ikx}}{H_{\Gamma}(e^{ik})}
  =
\int_{-\pi}^\pi \frac{dk}{2\pi}\, \frac{e^{ikx}}{i\Gamma - H(e^{ik})}
\, ,\end{equation}
where $\Gamma$ is the uniform dissipation motivated in Eq.~\eqref{eq:hamiltonian-gamma}. In this definition, we have replaced the discrete sum in Eq.~\eqref{eq:period} by an integral over continuous momenta, in this way taking the limit of infinite system size, $L\to\infty$.
As before, we reexpress the integral in Eq.~\eqref{eq:green} as a contour integral in order to use the residue theorem. In fact, there are two ways to do this:
\begin{equation}
\label{eq:green-contour}
  G(i\Gamma;x)
  =
  \frac1{2\pi i}
  \begin{dcases}
    \int\limits_{|z|=1} \,\frac{dz}z \frac{z^x}{i\Gamma - H(z)}\, , &
    \\
    \int\limits_{|\zeta|=1} \,\frac{d\zeta}\zeta \frac{\zeta^{-x}}{i\Gamma - H(\zeta^{-1})}
\, .&\end{dcases}
\end{equation}
Both integrals give the same result, but the first form is suitable for positive $x$, while the second is suitable for negative $x$. In both cases, the powers $z^x$, resp.\ $\zeta^{-x}$ do not contribute any poles inside the integrate contour (unit circle).
Again, we perform a partial fraction decomposition, and find that Eq.~\eqref{eq:partial-fraction} generalizes to
\begin{equation}
\label{eq:partial-inner}
    -\frac1{z}\frac1{i\Gamma-H(z)}
    = \sum_{j=1}^{2N}  G_j(\Gamma) \frac{1}{z-z_j(\Gamma)}
\, ,\end{equation}
where the roots $z_j(\Gamma)$ and coefficient matrices now depend on the uniform dissipation $\Gamma$. We keep the previous notation for the roots: $z_j$ is equivalent to $z_j(0)$.
For this formula to hold, we need to generalize our previous assumption (i) and now assume that the roots are distinct for each value $\Gamma \geq 0$. This also allows us to consider the values $z_j(\Gamma)$ as continuous functions in $\Gamma$, by ordering the roots appropriately.
Applying the residue theorem to Eq.~\eqref{eq:green-contour} and using the representation Eq.~\eqref{eq:partial-inner}, we obtain a generalization of Eq.~\eqref{eq:period-limit}:
\begin{equation}
\label{eq:green-explicit}
    G(i\Gamma;x) =
    \begin{cases}
        \sum\limits_{j\in J_R(\Gamma)} G_j(\Gamma) (z_j(\Gamma))^{x} \ ,
        &\!\!\!\!\text{if } x=0,1,\dots  \\
        \sum\limits_{j\in J_L(\Gamma)} (-1)G_j(\Gamma) (z_j(\Gamma))^{x}\ ,
        &\!\!\!\!\text{if } x=0,-1,\dots
.\end{cases}\end{equation}
Here, we have introduced a notation for the indices of the roots that occur in each sum:
\begin{subequations}
\label{eq:green-indices}
\begin{align}
  J_R(\Gamma) &:= \{j: |z_j(\Gamma)| < 1\}
\end{align}
records the indices of the roots which occur for positive $x$, while
\begin{align}
  J_L(\Gamma) &:= \{j: |z_j(\Gamma)| > 1\}
\end{align}
\end{subequations}
collects those that contribute for negative $x$.
We note that the roots are divided in such a way that the Green function decays in space: those with magnitude smaller than one are taken to a positive power, while the others are taken to a negative power, and the terms in the sum become exponentially small as the distance to the origin increases.

We now define the \emph{bulk Green function}, which describes the physical response of a non-Hermitian system in the bulk regime introduced in Section \ref{sec:bulk}, via analytic continuation. We try to make the definition of the bulk Green function plausible here, but refer to Appendix~\ref{sec:schroedinger} for details on the connection between long-time limit and analytic continuation.
First, we note that the Green function $G(0;x)$ always decays in space, because the integral in its definition Eq.~\eqref{eq:green} is the $L\to\infty$ limit of the sum Eq.~\eqref{eq:period-limit}  defining the periodic Green function, which we have shown to decay in space by explicit computation, Eq.~\eqref{eq:period-limit}.
For this reason, in general the Green function $G(0;x)$ will not agree with the bulk response of a system with amplification, which may grow in space. However, we argue that by  adding an uniform dissipation $\Gamma_0$ larger than the maximum gain,  the corresponding Green function $G(i\Gamma_0;x)$ is physical.
Thus, it is plausible that the bulk response of a general non-Hermitian Hamiltonian is obtained by gradually removing the extra dissipation via analytic continuation, in this way obtaining the bulk Green function
\begin{equation}
  \label{eq:bulk}
  G_{\text{bulk}}(0;x) := G(i\Gamma_0 \to 0; x)
.\end{equation}
Here, the notation $i\Gamma_0 \to 0$ stands for analytic continuation, which is performed as follows: The Green function $G(i\Gamma,x)$, given by Eqs.~\eqref{eq:green-explicit} and \eqref{eq:green-indices}, is evaluated for a fixed position $x$ and then analytically continued in the variable $\Gamma$ along a straight line from $\Gamma_0$ to $0$.
Crucially, the index sets $J_{L/R}(\Gamma_0)$ are kept fixed in the process of analytic continuation, such that $G(i\Gamma_0 \to 0; x)$ is different from $G(0,x)$ directly obtained from Eq.~\eqref{eq:green}, letting $\Gamma_0 = 0 $ inside the intergral. We will discuss the rationale behind keeping the index set fixed under analytic continuation in the next paragraph.

\begin{figure}
\includegraphics{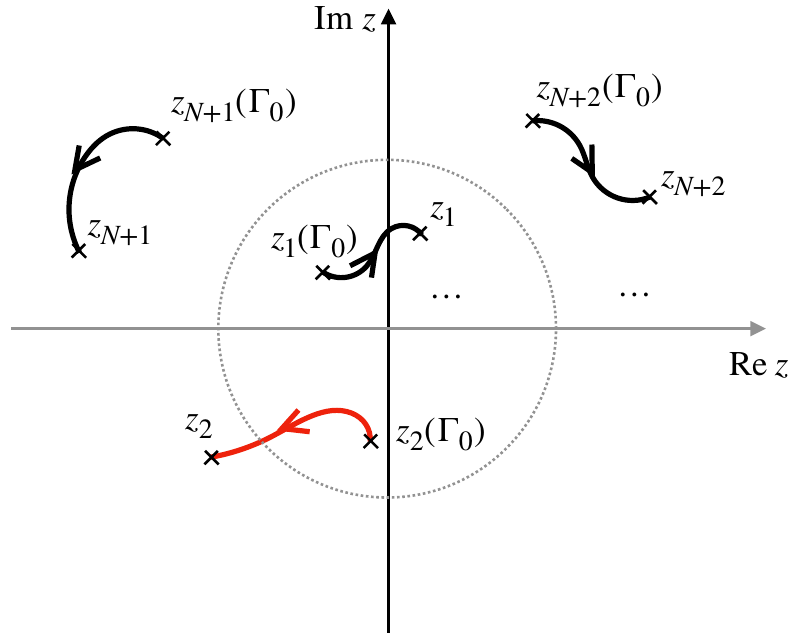}
\caption{%
Illustration of the analytic continuation $\Gamma_0 \to 0$. For large uniform dissipation, the roots $z_1(\Gamma_0),\dots,z_N(\Gamma_0)$ lie inside the unit circle, while the roots $z_{N+1}(\Gamma_0),\dots,z_{2N}(\Gamma_0)$ lie outside.
As the dissipation is reduced, the roots $z_j(\Gamma)$ will move along paths in the complex plane.
If one of the roots moves from inside the unit circle to the outside (or vice versa), say $z_2(\Gamma)$ as shown here, the bulk Green function $G_{\text{bulk}}$ will grow in space.
No such crossing may occur if the Hamiltonian is purely dissipative or has a real line gap.
\label{fig:roots_analytic}}%
\end{figure}

We now discuss the recipe for analytic continuation of the Green function $G(i\Gamma_0,x)$ in detail. First, one needs to realize that it does not work to perform the analytic continuation $i\Gamma_0 \to 0$ inside the integral Eq.~\eqref{eq:green}. In general, we will obtain different results depending on whether we first evaluate the integral, and then perform an analytic continuation, or whether we first perform an analytic continuation of the integrand and then integrate. Only in the case where the integrand $[i\Gamma-H(e^{ik})]^{-1}$ is analytic on the line $0\leq \Gamma < \Gamma_0$, both procedures will give the same result, and the periodic and the bulk Green functions agree with each other.
This condition of no poles on the upper complex axis is satisfied when the eigenvalues of the Bloch Hamiltonian avoid not just zero energy ($E=0$), but the upper imaginary axis ($E=i\Gamma$ for any $\Gamma \geq 0$). This certainly holds for purely dissipative Bloch Hamiltonians, whose eigenvalues avoid the entire upper half-plane. But it is also true for Bloch Hamiltonians with a real line gap as defined in the literature~\cite{Kawabata:2019b} and discussed earlier, because there the entire imaginary axis is avoided, though this is stronger than what we actually need here.
In contrast, in the case where the integrand becomes singular, the analytic continuation of the integrand will yield a different result than a continuation after performing the integral.
Thus, in general we have to define the analytic continuation via following the evolution of the zeros in Eqs.~\eqref{eq:green-explicit} and \eqref{eq:green-indices}.
When $\Gamma$ changes, the roots $z_j(\Gamma)$ evolve along paths in the complex plane, beginning at $z_j(\Gamma_0)$ and ending at $z_j=z_j(0)$ [see Fig.~\ref{fig:roots_analytic}]; similarly for the prefactors $G_j(\Gamma)$. Now, the key point is that the index sets $J_{L,R}(\Gamma)$ may evolve discontinuously, since one root can jump from one set to the other. However, the analytic continuation of the Green function must be continuous, and thus jumps in the index set are not allowed. For this reason, \emph{we keep the index set unchanged in the process of analytic continuation, and thus equal to $J_{L/R}(\Gamma_0)$.} In other words, the assignment of the roots to each sum is determined by their values for large dissipation, while the roots themselves are the same as for zero dissipation; we conclude that
\begin{equation}
\label{eq:bulk-explicit}
    G_{\text{bulk}}(0;x) =
    \begin{cases}
        \sum\limits_{j\in J_R(\Gamma_0)} G_j (z_j)^{x},
        &\!\!\!\text{if } x=0,1,\dots  ,\\
        \sum\limits_{j \in J_L(\Gamma_0)} (-1)G_j (z_j)^{x},
        &\!\!\!\text{if } x=0,-1,\dots
.\end{cases}\end{equation}
This result holds when our assumption that the roots $z_j(\Gamma)$ remain distinct along the path is met.
\footnote{For Eq.~\eqref{eq:bulk-explicit} to hold, the assumption can be relaxed: The roots $z_j(\Gamma)$ may coincide for some values of $\Gamma$ as long as they do not do so on the unit circle, i.e.\ simultaneously when a root jumps between the sets $J_{L,R}(\Gamma)$. This case is very special, but examples can be constructed, e.g.\ for special values of complex $t$ for the Hamiltonian discussed in Appendix \ref{sec:green-growth}.}
The representation of the bulk Green function Eq.~\eqref{eq:bulk-explicit} implies that the bulk Green function will grow exponentially in space if a root that was inside the unit circle for large dissipation has moved to the outside, and vice versa, because then a value with $|z_j|>1$ ($|z_j|<1$) is taken to a positive (negative) power $x$. Due to its exponential growth, the bulk Green function cannot agree with the periodic Green function in such a case, as the periodic Green function always decays in space.

This concludes our discussion of the periodic and the bulk Green function. The fact that the latter may grow in space is unique to non-Hermitian systems, and is not possible in the Hermitian case. We have derived an explicit criterion for this to happen, namely that a root $z_j(\Gamma)$ has moved from one side of the unit circle to the other during analytic continuation. In the next section, we will connect the possible motion of roots into or out of the unit circle to the non-Hermitian winding number.

%
\section{Non-Hermitian winding number}
\label{sec:winding}

In this section, we relate the non-Hermitian winding number, which is a topological invariant for periodic boundary conditions, to the bulk Green function. In particular, we show that if the non-Hermitian winding number is nonzero, then the bulk Green function grows in space, which may only happen in systems whose Hamiltonian is far from Hermitian.

By definition, the non-Hermitian winding number $\nu(H)$ counts how often the complex phase of the determinant $\det(H(z))$ winds around the origin as the Bloch momentum traverses the Brillouin zone:
\begin{equation}
  \label{eq:winding}
    \nu (H) := \int_{-\pi}^\pi \frac{dk}{2\pi}\, \frac{d}{dk}[\arg\det(H(e^{ik}))] \ 
.\end{equation}
Since the determinant of the Hamiltonian is the product of its eigenvalues, we see that this definition is viable only if the eigenvalues of the Bloch Hamiltonian are nonzero, i.e.\ if it has a point gap at the origin. Moreover, since the eigenvalues of a Hermitian Hamiltonian are real, we see that this invariant is always zero for Hermitian systems. Thus, a nonzero winding number identifies non-Hermitian Hamiltonians that lie outside the topological classification of Hermitian systems.

We can relate the winding number directly to the roots $z_j$ defined in the previous section, thanks to the factorization Eq.~\eqref{eq:determinant-factorization}. In particular, we can replace the phase ($\arg$) in the definition Eq.~\eqref{eq:winding} by a logarithm, and then express the integral Eq.~\eqref{eq:winding} as a contour integral over a logarithmic derivative
\begin{equation}
\label{eq:winding-contour}
    \nu (H) = \frac1{2\pi i}\oint_{|z|=1} \frac{\frac{d}{dz} \det(H(z))}{\det(H(z))} \,dz
.\end{equation}
With the help of Eq.~\eqref{eq:determinant-factorization}, the logarithmic derivative is readily evaluated to be equal to $-\frac{N}{z} + \sum_{j=1}^{2N} \frac1{z-z_j}$, and using the residue theorem, we find that the winding number is equal to $\nu(H) = -N + \sum_{|z_j|<1} 1$. Since the total number of roots is equal to $2N$, we find that the winding number is one half times the difference of the number of roots inside and outside the unit circle
\begin{align}
  \label{eq:winding-roots}
  \nu (H)
  &=
  \frac{1}{2} \left[\sum_{|z_j| < 1} 1 - \sum_{|z_j| > 1} 1\right]
.\end{align}

We can now argue that in the case of a nonzero winding number, $\nu(H) \neq 0$, the bulk Green function necessarily grows in space:
In the last section, we have seen that the bulk Green function grows in space if the location of a given root $z_j(\Gamma)$ changes in the complex plane between the inside and the outside of the unit circle. [see Fig.~\ref{fig:roots_analytic}]. For each Hamiltonian $H(\Gamma)$, we can consider the winding number $\nu(H_\Gamma)$, which counts the imbalance Eq.~\eqref{eq:winding-roots} of roots for each value of $\Gamma$.
The imbalance $\nu(H_\Gamma)$ changes by $\pm 1$ each time that a root crosses the unit circle; and only at these times. Specifically, the change is $+1$ when a root crosses from the outside to the inside, and $-1$ in the other direction. This means that if the starting and ending point of the continuation have differen winding numbers, $\nu (H) \neq \nu (H_{\Gamma_0})$, then a nonzero net number of roots must have has moved into or out of the unit circle to the other. But the explicit expression~\eqref{eq:green-explicit} shows that the Green function grows in space precisely when at least one root $z_j(\Gamma)$ has done so. Thus, we have found that the inequality of winding numbers is a sufficient criterion for spatial growth. (In case of equality, some roots may still have moved, we just cannot tell from the winding numbers alone.) If we can additionally show that the winding number for large dissipation is zero, then the claim follows.

\begin{figure}
\includegraphics{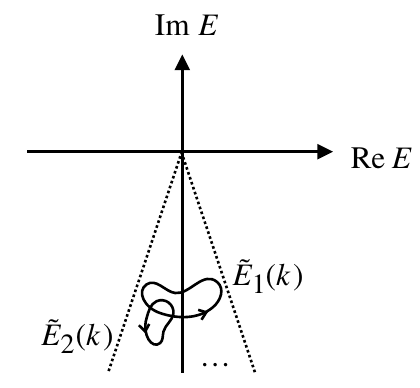}
\caption{%
Illustration of the paths of the eigenvalues of a non-Hermitian Bloch Hamiltonian. For large dissipation, they are contained within a small angle that enclosed the negative imaginary axis.
\label{fig-dissipative-winding}}%
\end{figure}

Thus, to conclude the argument, we need to show that  $\nu (H_{\Gamma_0})=0$. By definition, the dissipation $\Gamma_0$ is chosen such that all eigenvalues of the Hamiltonian $H_{\Gamma_0}(e^{ik})$ have negative imaginary part for all Bloch momenta $k$. For the proof, we consider an even larger dissipation $\Gamma=\lambda \Gamma_0$ with $\lambda \gg 1$.
Let $E_1(k),\dots ,E_N(k)$ denote the eigenvalues of the original Bloch Hamiltonian $H(e^{ik})$.
If the dissipation is sufficiently large, the eigenvalues $\tilde{E}_{n}(k) = E_n(k) - i\lambda\Gamma_0 $ of the Hamiltonian $H_{\lambda\Gamma_0}$ will be contained within a small angle that encloses the negative imaginary axis, $|\Re \tilde{E}_n(k)| \ll -\Im \tilde{E}_n(k)/(2N)$ uniformly in $k$. The winding number of the determinant is the sum of the winding angles of the individual eigenvalues (which can interchange). This is at most $N$ times this small angle, which is smaller than a full rotation [see Fig.~\ref{fig-dissipative-winding}]. Thus, the winding number for very large dissipation vanishes, $\nu(H_{\lambda\Gamma_0}) = 0$. Now, gradually removing the extra dissipation by sending $\lambda\to 1$ will not change the winding number, because the eigenvalues have a negative imaginary part as long as $\Gamma > \Gamma_0$, and their product, the determinant $\det(H_{\lambda\Gamma_0}(e^{ik}))$, cannot be equal to zero. Thus, the winding numbers agree, $\nu(H_{\Gamma_0}) = \nu(H_{\lambda\Gamma_0}) = 0$ for any $\lambda > 1$, and this concludes the argument.

%
\section{Conclusion}
\label{sec:conclusion}

In this work, we have defined and studied the response of a non-Hermitian system in the bulk regime, by which we mean the response to a perturbation on a timescale where the induced excitation has not propagated to the boundary yet, so that boundary conditions do not yet influence the response. We have shown that the bulk response grows exponentially in space in the case where the non-Hermitian winding number is nonzero. This spatial growth is due to amplification, which is not present in  Hermitian systems and implies a high sensitivity to boundary conditions; therefore it is a key mechanism for the breakdown of the bulk-boundary correspondence, where topological invariants calculated for periodic boundary conditions no longer fully capture the system response for open boundary conditions. We also have shown that a spatial growth does not occur for systems with a real line gap, which suggests that the bulk-boundary correspondence established for Hermitians systems still holds in this case.

To find the response in the bulk regime, we considered the Green function for periodic boundary conditions, but could not straightforwardly take the limit of the system size going to infinity, as this limit entails that excitations propagate through the entire system repeatedly. Instead, we have first identified the subclass of purely dissipative non-Hermitian Hamiltonians, characterized by the fact that all eigenvalues of the Bloch Hamiltonian lie in the lower complex plane, and argued that the bulk response does agree with the periodic Green function in this case. To find the bulk Green function in the general case, we have added a uniform dissipation $-i\Gamma$ to the Hamiltonian $H(z)$, and argued that since the system becomes purely dissipative when the dissipation is larger than the maximum gain $\Gamma_0$, we can obtain the response at zero dissipation by performing an analytic continuation $i\Gamma_0\to 0$, pointwise for every lattice site. We could show that the periodic and bulk Green function still agree in the case where the Bloch Hamiltonian has a real line gap, by using that the analytic continuation of an integral is related to the analytic continuation of the integrand. For the general case, we were able to derive explicit formulas for the Green functions in terms of the complex roots $z_j(\Gamma)$, which are the zeros of the determinant $\det(H(z)-i\Gamma)$. This enabled us to argue that the bulk Green function grows in space if at least one of the roots moves from inside the unit circle to the outside or vice versa. But at the same time, we could relate the roots to the non-Hermitian winding number, which is one half times the difference of the number of roots inside and outside the unit circle, and thus show that it detects when a root crosses the unit circle in this way.

Our results suggest that the physics of non-Hermitian systems becomes richer than that of Hermitian systems only in a regime with strong gain. However, in this regime, basic assumptions have to be revisited; as we have shown, the response in the bulk regime and the response for a periodic system in the limit of large system size no longer need to agree with each other. In contrast, we conjecture that purely dissipative systems behave very similarly to Hermitian systems.

%
\section*{Acknowledgements}
We would like to thank G.~Refael for helpful discussions on the content of this manuscript.
B.~R.\ and H.-G.~Z.\ acknowledge financial support from DFG Grant RO 2247/11-1. 

\appendix

%
\section{General hopping matrices}
\label{sec:general-hopping}

In this appendix, we extend our discussion of the bulk Green function and the non-Hermitian winding number to the case where the hopping matrices $H_{-1}$ and $H_1$ in the Hamiltonian \eqref{eq:hamiltonian} are not assumed to be invertible.

This case is relevant because an often discussed non-Hermitian variant of the Su-Schrieffer-Heeger (SSH) model~\cite{Lee:2016,Kunst:2018,Yao:2018b,Xiong:2018,Lee:2019c,Herviou:2019} has hopping matrices
\begin{multline}
  H_{\text{SSH}}(z) =
  \begin{pmatrix}
    0 & t_2 \\
    0 & 0
  \end{pmatrix}
  z^{-1}
  +
  \begin{pmatrix}
    0 & t_1+\gamma/2 \\
    t_1 - \gamma/2 & 0
  \end{pmatrix}
  \\ +
  \begin{pmatrix}
     0 & 0 \\
     t_2 & 0
  \end{pmatrix}
  z
,\end{multline}
which have rank one instead of two. More generally, this case will occur in any model with nearest-neighbor hopping where the hopping between unit cells only involves a small number of sites. Fortunately, the gist of our argument remains unchanged.

The main difference is that we now have fewer than $2N$ roots, and the determinant \eqref{eq:determinant-polynomial} is modified to
\begin{equation}
    \det(zH(z)) = \sum_{m=M_-}^{M_+} a_m z^m
.\end{equation}
Here, $M_-$ ($M_+$) is the lowest (highest) index such that the complex coefficient $a_{M_-}$ and ($a_{M_+}$) is nonzero.
This polynomial has $M=M_+-M_-$ nonzero roots $z_1,\dots,z_M$.
Since the outermost coefficients are given by the determinants of the hopping matrices, $a_0=\det(H_{-1})$ and $a_{2N}=\det(H_1)$, we are assured that the number of roots is smaller than $2N$ if these matrices are not invertible.
The factorization Eq.~\eqref{eq:determinant-factorization} becomes
\begin{equation}
  \label{eq:determinant-factorization-general}
  \det(H(z)) = a_{M_+} z^{M_- -N} (z-z_1)\cdots(z-z_M)
.\end{equation}
In the partial fraction decomposition Eq.~\eqref{eq:partial-fraction}, we now have to take into account that there will be additional poles at $z=0$ and $z=\infty$, possibly of higher order, which give rise to an additional Laurent polynomial:
\begin{equation}
  -\frac1{z} \frac{1}{H(z)}
  =
  \sum_{j=1}^{M} G_j \frac1{z-z_j}
  + \sum_{m=m_{\text{min}}}^{m_{\text{max}}} b_m z^m
,\end{equation}
where the $b_m$ are complex coefficients. We do not attempt to determine the order of these poles more precisely, but from the adjugate matrix, Eq.~\eqref{eq:adjugate}, we can estimate that $m_{\text{min}} \geq -M_-$, and $m_{\text{max}} \leq 2(N-1)-M_+$.
Fortunately, the effect of these poles on the explicit expressions for the Green functions, Eqs.~\eqref{eq:period-explicit}, \eqref{eq:period-limit}, \eqref{eq:green-explicit}, and \eqref{eq:bulk-explicit}, is rather benign: Since the integrand in Eqs.~\eqref{eq:period-contour} and \eqref{eq:green-contour} involves a power $z^x$ or $\zeta^{-x}$, we see that these poles are removed whenever the position of $x$ is sufficiently far from the center $x=0$, and, in the periodic case, also sufficiently far away from $x=L$. Here, ``sufficiently far'' means that the magnitude of the spatial distance $x$ has to be larger than $m_{\text{max}}$ and $m_{\text{min}}$. In other words, the shape of the Green functions is altered in the vicinity of $x=0,L$, but the explicit expressions are unchanged at larger distances. This means that our arguments in Sec.~\ref{sec:bulk-green} about analytic continuation and exponential growth are mostly unchanged.

For the non-Hermitian winding number, using the roots with the contour integral \eqref{eq:winding-contour}, we find that
\begin{subequations}
  \begin{align}
    \nu (H) &= (M_- - N) + \sum_{|z_j|<1} 1
    \\ &=
    (M_+ - N) - \sum_{|z_j| > 1} 1
  ,\end{align}
\end{subequations}
where we have used that no root lies on the unit circle, and the total number of roots is $M = M_+ - M_-$. While the winding number no longer counts the imbalance of roots inside and outside the unit circle, the argument in Sec.~\ref{sec:winding} remains essentially unchanged: The winding number still changes by $\pm 1$ precisely when one of the roots $z_j(\Gamma)$ crosses the unit circle, and we still have that $\nu(H_{\Gamma_0}) = 0$ for large dissipation $\Gamma_0$. Thus, as before, we conclude that a nonzero non-Hermitian winding number implies that the bulk Green function grows in space.

%
\section{Bulk regime: Long-time limit equals analytic continuation}
\label{sec:schroedinger}

In this appendix, we justify the definition of the bulk Green function as the analytic continuation Eq.~\eqref{eq:bulk} by showing that it agrees with the solution to the driven Schr\"{o}dinger equation Eq.~\eqref{eq:driven-schroedinger} in the bulk regime whenever the long-time limit of the wave function exists in a pointwise sense. We show this by using the Laplace transform technique. Our discussion applies to lattice systems with translation symmetry.

First, we recapitulate the definition of a Green function as an operator inverse, and discuss its uniqueness for an infinite system size. A Green function $G(E;x,y)$ in the energy domain is defined to be a solution to the equation $(E-\hat H)G(E;x,y) = \Id \delta_{xy}$, where $\hat H$ is the Hamiltonian operator, acting on the variable $x$. Such a solution exists whenever the energy $E$ is not in the spectrum of the operator.
For finite systems, the Green function is unique, but for infinite systems, there is always more than one Green function, because in this case, there exists a multitude of wave functions $\psi_{E}(x)$ that solve the homogeneous equation, $(E-\hat H)\psi_{E} = 0$. To restore uniqueness, one has to impose boundary conditions, for example that the Green function vanishes as $x \to \pm \infty$. In fact, this is precisely how the Green function Eq.~\eqref{eq:green} can be understood: It is the operator inverse that decays at infinity. For a generic gapped Hamiltonian with translation symmetry, this makes it unique: The difference of two Green functions is a solution to the homogeneous equation, which generically grows in at least one spatial direction, because such solutions are linear combination of functions of the form $z_j^{x}$ where $z_j$ are solutions to the equation $\det(E-H(z_j))=0$. One major exception to this rule are Hamiltonians with flat bands, which are characterized by eigenstates that are localized on a few lattice sites; we will ignore this case and focus on generic Hamiltonians.

To connect the Schr\"{o}dinger equation Eq.~\eqref{eq:driven-schroedinger} to the Green function, we use the Laplace transform. In Hermitian systems, one would use the Fourier transform, but in non-Hermitian systems, we have to consider that the wave function may grow in time, and the Fourier integral may not converge. The Laplace transform of a function $\tilde{\psi}(t,x)$ that grows at most exponentially in time is given by
\begin{equation}
\label{eq:laplace}
    L[\tilde{\psi }](\Gamma)
    :=
    \int _0^\infty \tilde{\psi }(t,x) e^{-\Gamma t}dt
\ .\end{equation}
This integral converges absolutely and represents a complex analytic function in the variable $\Gamma$ in the region $\Re \Gamma > \Gamma_0$, where $\Gamma_0$ is the maximal growth rate of the wave function, that is $|\tilde{\psi}(t,x)|$ grows asymptotically no faster than the exponential $e^{\Gamma_0 t}$. Representing the wave function in the driven Schr\"{o}dinger equation Eq.~\eqref{eq:driven-schroedinger} in this way, the equation is transformed into
\begin{align}
    (E + i\Gamma - \hat H) L[\tilde{\psi }](\Gamma) &= \Gamma^{-1} \psi_0\delta_{x,y}
\ ,\end{align}
where we have absorbed the oscillation in time by setting $\tilde{\psi }(t,x) = e^{iEt}\psi (t,x)$ and have positioned the source at point $y$.
The above equation implies that the Laplace transform yields an operator inverse, and the expression
\begin{equation}
  \label{eq:green-laplace}
  \tilde{G}_{\sigma_1 \sigma_2}(E+i\Gamma;x,y) := L[\partial_t\psi_{(\sigma_2)}](\Gamma)_{\sigma_1}
\end{equation}
defines a Green function in the sense above. For clarity, we have made explicit the indices $\sigma_1,\sigma_2=1,\dots,N$ for the internal degrees of freedom on each lattice site, highlighting that $G$ is a matrix-valued function of the positions $x$ and $y$. On the right-hand side, the subscript $\sigma_1$ refers to the component of the vector, while the notation $\psi_{(\sigma_2)}$ refers to the solution computed for a source vector with components given by $(\psi_0)_\sigma = \delta_{\sigma_2 \sigma}$; in the following, we will suppress theses indices again.
For the definition above, we have removed a factor of $\Gamma$ by using that the Laplace transform of the time derivative satisfies $L[\partial _t \tilde{\psi }](\Gamma) = \Gamma L[\tilde{\psi }](\Gamma) - \tilde{\psi}(0,x)$.
It is not a priori clear that this Green function is equal to the bulk Green function defined in the main text, Eq.~\eqref{eq:bulk}, because the operator inverse of an infinite system is not unique, as we have pointed out in the beginning of this appendix.
However, we will now argue that in the region $\Gamma > \Gamma_0$, where the Laplace transform is guaranteed to converge absolutely, we have
\begin{equation}
  \tilde{G}(E+i\Gamma;x,y) = G(E+i\Gamma;x-y)
  \quad\text{ for } \Gamma > \Gamma_0
,\end{equation}
where $G$ is the Green function defined by spatial Fourier transform in Eq.~\eqref{eq:green} (for general $E$).
To see this, it suffices to show that both sides of the equation satisfy the same boundary conditions, because, as we have discussed in the beginning, the operator inverse becomes unique for generic Hamiltonians once boundary conditions are imposed.
The Green function $G$ vanishes at large distances $x\to \pm\infty$ by definition.
The Green function $\tilde{G}$ also vanishes, because it can be interpreted as a solution to the driven Schr\"{o}dinger for the Hamiltonian $\hat H_{\Gamma} = \hat H - i\Gamma$. Since $\Gamma > \Gamma_0$, this Hamiltonian is purely dissipative, and it is physically plausible that excitations decay as they travel away from the position of the drive. For a mathematical justification, we refer to Appendix~\ref{sec:green-decay}. 
We note that for a Hermitian system, any dissipation $\Gamma$, no matter how small, will give rise to a purely dissipative Hamiltonian, and this explains why in this case, the bulk Green function is given by the Fourier transform with the $E+i0^+$ description.

We next show that if the long-time limit $t\to \infty$ exists, then it is equal to the bulk Green function Eq.~\eqref{eq:bulk} defined by analytic continuation. The key point is that if there is a stationary state, then it can be expressed as an integral over the time derivative, and this integral is the Laplace transform at $\Gamma=0$:
\begin{equation}
  \label{eq:limit-t-infty}
    \lim_{t\to \infty } \tilde{\psi }(t,x) = \int _0^\infty \partial _t \tilde{\psi }(t,x) \,dt = L[\partial _t \tilde{\psi }](0)
.\end{equation}
The sense in which the limit is taken and in which this integral converges is important. Here, we assume pointwise convergence, i.e.\ for each position $x$, we ask that the above integral over the time derivative converges absolutely, which implies that at each position, the wave function converges to a stationary state as $t\to\infty$. Then, the absolute convergence of the integral Eq.~\eqref{eq:laplace} defining $L[\partial_t\tilde{\psi}](\Gamma)$ is ensured for any $\Re \Gamma>0$, and the analytic continuation along the straight line from $\Gamma_0$ to $0$ just gives the value at $\Gamma=0$. Thus, the stationary state agrees with the continuation that we have described in the main text, where we evaluated Eq.~\eqref{eq:green} for a fixed position $x$.
In contrast, had we insisted on convergence in the Hilbert space norm, then the existence of the limit would entail that the norm $\norm{\psi(t)}$, obtained by integrating over position $x$, stays finite as $t\to \infty$. This is unsuitable for describing a stationary state that grows exponentially in space.

We have already mentioned in the main text that convergence to a stationary state is not always guaranteed. Intuitively, it requires that the non-Hermitian Hamiltonian has the property that excitations travel faster in space than they grow in time. In this case, the wave function has a chance to converge at a position $x$, as excitation travel away fast enough, but these excitations will contribute to positions further away. The motivating example Eq.~\eqref{eq:continuum-long-time} is of this form. However, for some systems, it may happen that the wave function grows in time even at a single site, so that the limit \eqref{eq:limit-t-infty} does not exist even in the pointwise sense. However, the analytic continuation Eq.~\eqref{eq:bulk} will still give a result, though it may be unintuitive. Determining precise conditions for the existence of a stationary state as a pointwise limit is beyond the scope of this work, but we present an explicitly solvable example in Appendix~\ref{sec:long-times} where both existence and absence occur for different parameter regimes.

%
\section{Decay of bulk Green function for purely dissipative Hamiltonians}
\label{sec:green-decay}

In this appendix, we provide a mathematical argument that the Laplace transform Eq.~\eqref{eq:green-laplace}, which gives rise to the bulk Green function, vanishes as $x\to \pm \infty $ if the parameter $\Gamma $ is large enough so that the translationally invariant Hamiltonian with added dissipation, $\hat H - i\Gamma $, is purely dissipative.

To recapitulate, we consider the solution $\psi(t,x)$ to the initial value problem
\begin{subequations}
\begin{align}
    i\partial _t \psi (t,x)
    &= \hat H \psi (t,x) - \psi _0e^{-iEt}\delta_{x,y}
    \\
    \psi (0,x) &= 0
.\end{align}
\end{subequations}
As before, we absorb the oscillatory time dependence and define $\tilde{\psi }(t,x) = e^{iEt}\psi (t,x)$.
Our goal is to show that, as a function of $x$,
\begin{equation}
    \label{eq:laplace-decay}
    L[\tilde{\psi}](\Gamma ) \equiv \int _0^\infty \tilde{\psi }(t,x)e^{-\Gamma t} \, dt
    \to 0
    \text{ for } x\to \pm \infty 
\end{equation}
for large enough $\Gamma > 0$.
To establish this, we will argue that the Hilbert space norm of the Laplace transform is finite, $\norm{L[\tilde{\psi}](\Gamma)} < \infty$. Since the norm is an infinite sum over position $x$, this will indeed imply that the above expression vanishes as $x \to \pm\infty$, because otherwise the sum would be infinite.

To establish the finiteness of the norm of the time integral in Eq.~\eqref{eq:laplace-decay}, we estimate the norm of the integrand. But since the time integral extends to infinity, it would not be enough to estimate the integrand by a constant; instead, we will establish an estimate with an exponential decay,
\begin{equation}
  \label{eq:estimate-exp}
  \norm{\tilde\psi(t)}e^{-\Gamma t} \leq C e^{-\eta t}
,\end{equation}
for some constant $C>0$ and some small $\eta > 0$ which is to be determined. Then, we can estimate
\begin{equation}
  \norm{L[\tilde{\psi}](\Gamma)}
  \leq \int_0^\infty \norm{\tilde\psi(t)}e^{-\Gamma t}\, dt
  \leq C \int_0^\infty e^{-\eta t}\, dt
  \leq \frac{C}{\eta}
,\end{equation}
which is finite, and the result follows.
To show Eq.~\eqref{eq:estimate-exp}, we introduce a rescaled wave function $\phi (t,x) = e^{-(\Gamma-\eta)t}\tilde{\psi }(t,x)$, so that our goal becomes showing that $\norm{\phi(t)}<C$. We observe that this wave function satisfies the modified Schr\"{o}dinger equation
\begin{subequations}
\label{eq-schroedinger-decay}
\begin{align}
    i\partial _t \phi &= \hat H_1 \phi - e^{-(\Gamma-\eta)t}\psi _0 \delta_{x,y}
    \\
    \phi (0,x) &= 0    
\end{align}
\end{subequations}
with Hamiltonian $\hat H_1 = \hat H - i(\Gamma-\eta) - E$. Since $\Gamma$ is larger than the maximum gain, the slightly smaller quantity $\Gamma-\eta$ will still be larger than the maximum gain if we just choose $\eta$ small enough. For example, if $\Gamma=\Gamma_0$ as in Eq.~\eqref{eq:gamma-maximum-gain}, we can choose $\eta=\varepsilon/2$.
With this choice, the Hamiltonian $\hat H_1$ is purely dissipative, which implies that the time evolution does not increase the Hilbert space norm
\begin{equation}
    \norm{e^{-it\hat H_1}\tilde{\phi }} \leq \norm{\tilde{\phi }}    
\label{eq-norm-dissipation}
\end{equation}
for any wave function $\tilde{\phi }(x)$. The fact that dissipation cannot increase the probability to the wave function is plausible, and we will prove it by employing Fourier transformation in the next paragraph. The solution to the driven Schr\"o{}dinger equation~\eqref{eq-schroedinger-decay} is explicitly found as
\begin{equation}
    \phi (t,x) = i\int _0^t d\tau \, e^{-i(t-\tau )\hat H_1}e^{-\tau (\Gamma-\eta)} \psi _0 \delta_{x,y}
\end{equation}
and we can use the estimate Eq.~\eqref{eq-norm-dissipation} to conclude that
\begin{equation}
    \norm{\phi (t)}
    \leq \int _0^t d\tau \, e^{-\tau (\Gamma-\eta)} \norm{\psi _0}
    \leq (\Gamma-\eta)^{-1}\norm{\psi _0}
.\end{equation}
This is the desired estimate.

To conclude the argument, we need to show that the time evolution with a purely dissipative Hamiltonian $\hat H_1$ does not increase the Hilbert space norm, Eq.~\eqref{eq-norm-dissipation}. To achieve this, we use the fact that any function $\tilde\phi(x)$ on the lattice with finite Hilbert space norm has a Fourier transform $\tilde{\phi}_k$ with real momenta $k$. By Parseval's identity, the Hilbert space norm can be expressed as an integral over the Fourier components
\begin{equation}
  \norm{\tilde \phi}^2 \equiv \sum_{x} \norm{\tilde\phi(x)}^2 = \int_{-\pi}^\pi \frac{dk}{2\pi}\, \norm{\tilde\phi_k}^2
.\end{equation}
Thus, it is sufficient to show that the norm of each Fourier component individually does not increase. 
Crucially, since the Hamiltonian is translationally invariant, the time evolution $\tilde\phi(t,x)\equiv e^{-it\hat H_1} \tilde\phi(0,x)$ becomes a matrix multiplication applied to each Fourier component individually, that is
\begin{equation}
  \tilde\phi_k(t) = e^{-itH_1(e^{ik})}\tilde\phi_k(0)
,\end{equation}
where $H_1(e^{ik})$ is the Bloch Hamiltonian, which is an $N\times N$ matrix for each momentum. To show that the norm $\norm{\phi_k(t)}^2$ does not increase, we compute its time derivative. To simplify notation, we abbreviate  $\tilde{H} := H_1(e^{ik})$ and obtain
\begin{subequations}
\begin{align}
  \partial_t \norm{\phi_k(t)}^2
&=
  \<\partial_t \phi_k | \phi_k\> + \< \phi_k | \partial_t \phi_k \>
\\ &=
  i \< i\partial_t \phi_k | \phi_k\> - i \< \phi_k | i\partial_t \phi_k \>
\\ &=
  i \< \tilde{H} \phi_k | \phi_k\> - i \< \phi_k | \tilde{H} \phi_k \>
\\ &=
  i \< \phi_k | \tilde{H}^\dagger \phi_k\> - i \< \phi_k | \tilde{H} \phi_k \>
\\ &=
  \< \phi_k | i(\tilde{H}^\dagger - \tilde{H}) \phi_k \>
= \< \phi_k | 2\tilde{\Gamma} \phi_k \>
\label{eq:norm-decrease}
,\end{align}
\end{subequations}
where we have introduced the matrix $\tilde{\Gamma} = (i/2)(\tilde{H}^\dagger - \tilde{H})$. By definition Eq.~\eqref{eq:purely-dissipative}, the Hamiltonian $\hat H_1$ is purely dissipative if this matrix is negative semidefinite for any momentum, $\tilde{\Gamma} \leq 0$, which implies that the right-hand side of the last equation above is always nonpositive. Thus, the norm does not increase during time evolution, as claimed.

%
\section{Explicit solution to the driven Schr\"{o}dinger equation}
\label{sec:long-times}

In this appendix, we describe an explicit solution to the driven Schr\"{o}dinger equation for the Hamiltonian $H(z) = a - z$ where $a$ is a complex parameter. This Hamiltonian acts on one-component wave functions. As in the discussion of Eq.~\eqref{eq:hamiltonian}, the complex variable $z$ generalizes the phase $e^{ik}$ of the Bloch Hamiltonian. This system is decidedly non-Hermitian since it only includes hopping to the left, but not to the right, and can thus be thought of as a lattice analogue of the Hamiltonian Eq.~\eqref{eq:continuum-long-time}, which describes a wave that travels in one direction only.
We include this example here, because it provides insight into the existence of the long-time limit in a pointwise sense, as introduced in Section~\ref{sec:bulk} and Appendix~\ref{sec:schroedinger}. Determining the existence of this limit for general Hamiltonians is outside the scope of this work however.

In fact, for any general lattice Hamiltonian $H(z)$, the solution to the driven Schr\"{o}dinger equation at energy $E=0$,
\begin{subequations}
\begin{align}
    i\partial _t \psi(t,x)
    &= \hat H \psi(t,x) - \psi_0\delta_{x,y}
    \\
    \psi(0,x) &= 0
,\end{align}
\end{subequations}
can be obtained as a contour integral
\begin{equation}
    \psi(t,x) = \frac1{2\pi i} \oint_{|z|=1}[1 - e^{-itH(z)}]\psi_0 \frac1{H(z)} z^{x-y} \frac{dz}z
.\end{equation}
To see this, one directly checks that this formula satisfies both equations. Taking the time derivative yields the slightly simpler formula
\begin{equation}
    i\partial_t \psi(t,x) = -\frac1{2\pi i} \oint_{|z|=1} e^{-itH(z)}\psi_0 z^{x-y} \frac{dz}z
\label{eq-derivative-schroedinger}
.\end{equation}

We now specialize to $H(z) = a-z$. Without loss of generality, we consider a source at $y=0$ with strength $\psi_0=1$. Then, expanding the exponential as a power series in $z$,
\begin{equation}
    e^{-itH(z)} = e^{-it(a-z)} = e^{-ita} \sum_{k=0} \frac{(itz)^k}{k!}
,\end{equation}
and applying the residue theorem in Eq.~\eqref{eq-derivative-schroedinger}, we find that the contour integral picks out individual coefficients of the series; we obtain
\begin{equation}
    i\partial _t \psi (t,x) = -e^{-it a} \begin{cases}
        \frac{(it)^k}{k!}\delta_{k,-x}, &\text{ if } x \leq 0 \\
        0, &\text{ if } x > 0
    .\end{cases}
\end{equation}
Using integration by parts, we can find the wave function on particular lattice sites:
\begin{subequations}
\begin{align}
    \psi (t,x=0)  &= a^{-1}[1 - e^{-ita}]
    \\
    \psi (t,x=-1) &= a^{-2}[1 - e^{-ita}(1 + ita)]
    \\
    \psi (t,x=-2) &= a^{-3}[1 - e^{-ita}(1 + (ita) + (ita)^2/2)]
.\end{align}
\end{subequations}
A pattern becomes apparent, and one can check directly that the full solution is
\begin{equation}
    \psi (t,x) = \begin{cases}
        a^{x-1} \left[1 - e^{-ita} \sum_{k=0}^{-x}\left[\frac{(ita)^k}{k!}\right]\right], &\text{ if } x \leq 0 \\
        0, &\text{ if } x > 0
    .\end{cases}
\end{equation}
We can discuss the long-time limit of this expression: If the parameter has negative imaginary part, $\Im a < 0$, then the exponential vanishes as $t\to \infty$ and the solution converges pointwise to the stationary state
\begin{equation}
  \label{eq:example-stationary}
    \psi (t\to \infty,x) = \begin{cases}
        a^{x-1}, &\text{ if } x \leq 0 \\
        0, &\text{ if } x > 0
    .\end{cases}
\end{equation}
But when the parameter has positive imaginary part, the wave function diverges at every site. However, the analytic continuation still reproduces the wave function above, because it amounts to replacing $a \to a - i\Gamma$ with a large $\Gamma>0$ before taking the limit $t \to \infty$.

%
\section{A Hamiltonian whose bulk Green function grows}
\label{sec:green-growth}

In this appendix, we present a non-Hermitian Hamiltonian whose bulk Green function grows in space, but which can be deformed into a Hermitian Hamiltonian without closing the point gap. Since this implies that the non-Hermitian winding number of the Hamiltonian is zero, this provides an example that it is not correct to conclude from a vanishing winding number that the response would decay in space. This example also shows that the spatial growth of the Green function by itself is not topologically invariant under such deformations.

The Hamiltonian is derived from the non-Hermitian Su-Schrieffer-Heeger (SSH) model,~\cite{Lee:2016,Yao:2018b,Lieu:2018a,Esaki:2011} and given by
\begin{equation}
    H(z) = \begin{pmatrix}
        0 & it + z^{-1} \\
        it + z & 0
    \end{pmatrix}
\end{equation}
where $t>0$ is a positive real parameter which also satisfies $t\neq 1$. The transformation $t \to te^{i\phi}$ from $\phi=0$ to $\phi=-\pi/2$ is a homotopy to a gapped Hermitian Hamiltonian during which the eigenvalues never cross zero energy.
As explained in Section~\ref{sec:bulk-green}, the bulk Green function is obtained by computing the roots for a large dissipation and analytically continuing their location in the complex plane. The roots are the solutions to the quadratic equation $\det(H(z)-i\Gamma)=0$ and are given by
\begin{equation}
    z_{\pm}(\Gamma) = \frac{i}{2}\left[\frac{1+\Gamma^2 - t^2}{t} \pm \sqrt{\frac{(1+\Gamma^2-t^2)^2}{t^2}+4}\right]
.\end{equation}
For large $\Gamma$, we clearly have $|z_+(\Gamma)|>1$. The roots satisfy $z_+(\Gamma)z_-(\Gamma)=1$, which implies $|z_-(\Gamma)|<1$. Performing the analytic continuation to $\Gamma=0$, we find
\begin{equation}
    z_+(0) = it^{-1}
    ,\text{ and }
    z_-(0) = -it
.\end{equation}
Thus, from Eq.~\eqref{eq:bulk-explicit}, we see that the bulk Green function decays in space if $0<t<1$, but grows exponentially if $t>1$.

\bibliography{bib-non-hermitian}

\end{document}